\documentclass{article}
\usepackage{arxiv}

\usepackage[utf8]{inputenc} 
\usepackage[T1]{fontenc}    
\usepackage{hyperref}       
\usepackage{url}            
\usepackage{booktabs}       
\usepackage{amsfonts}       
\usepackage{nicefrac}       
\usepackage{microtype}      
\usepackage{lipsum}		
\usepackage{graphicx}
\usepackage{doi}
\usepackage{graphicx}%
\usepackage{multirow}%
\usepackage{amsmath,amssymb,amsfonts}%
\usepackage{amsthm}%
\usepackage{mathrsfs}%
\usepackage[title]{appendix}%
\usepackage{xcolor}%
\usepackage{textcomp}%
\usepackage{manyfoot}%
\usepackage{booktabs}%
\usepackage{algorithm}%
\usepackage{algorithmicx}%
\usepackage{algpseudocode}%
\usepackage{listings}%

\usepackage{amsmath}
\DeclareMathOperator*{\argmax}{argmax}

\theoremstyle{thmstyleone}%
\newtheorem{theorem}{Theorem}
\newtheorem{definition}{Definition}%
\newtheorem{corollary}{Corollary}%
\newtheorem{lemma}{Lemma}%
\newtheorem{proposition}{Proposition}%

\title{A greedy heuristic for graph burning}

\author{ \href{https://orcid.org/0000-0001-6334-2305}{\includegraphics[scale=0.06]{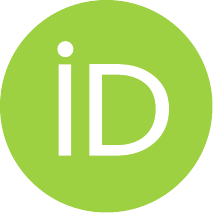}\hspace{1mm}Jesús García-Díaz}\\
	Coordinación de ciencias computacionales\\
	Instituto Nacional de Astrofísica, Óptica y Electrónica\\
	Puebla, Mexico 72840 \\
	\texttt{jesus.garcia@conahcyt.mx} \\
	\And
	\href{https://orcid.org/0000-0001-7168-2799}{\includegraphics[scale=0.06]{orcid.pdf}\hspace{1mm}José Alejandro Cornejo-Acosta} \\
	División de ingeniería informática\\
	Tecnólogico Nacional de México\\
	Guanajuato, Mexico 36425 \\
	\texttt{alejandro.ca@purisima.tecnm.mx} \\
	\And
	\href{https://orcid.org/0000-0001-9326-7713}{\includegraphics[scale=0.06]{orcid.pdf}\hspace{1mm}Joel Antonio Trejo-Sánchez} \\
	Centro de Investigación en Matemáticas\\
	Yucatán, Mexico 97302 \\
	\texttt{joel.trejo@cimat.mx} \\
}

\renewcommand{\shorttitle}{A greedy heuristic for graph burning}

\hypersetup{
	pdftitle={A template for the arxiv style},
	pdfsubject={q-bio.NC, q-bio.QM},
	pdfauthor={David S.~Hippocampus, Elias D.~Striatum},
	pdfkeywords={First keyword, Second keyword, More},
}

\begin{document}
	\maketitle
	
	\begin{abstract}
Given a graph $G$, the optimization version of the graph burning problem seeks for a sequence of vertices, $(u_1,u_2,...,u_p) \in V(G)^p$, with minimum $p$ and such that every $v \in V(G)$ has distance at most $p-i$ to some vertex $u_i$. The length $p$ of the optimal solution is known as the burning number and is denoted by $b(G)$, an invariant that helps quantify the graph's vulnerability to contagion. This paper explores the advantages and limitations of an $\mathcal{O}(mn + pn^2)$ deterministic greedy heuristic for this problem, where $n$ is the graph's order, $m$ is the graph's size, and $p$ is a guess on $b(G)$. This heuristic is based on the relationship between the graph burning problem and the clustered maximum coverage problem, and despite having limitations on paths and cycles, it found most of the optimal and best-known solutions of benchmark and synthetic graphs with up to 102400 vertices. Beyond practical advantages, our work unveils some of the fundamental aspects of graph burning: its relationship with a generalization of a classical coverage problem and compact integer programs. With this knowledge, better algorithms might be designed in the future.
	\end{abstract}


	\section{Introduction}\label{sec1}
	
\begin{figure}[h]
	\centering
	\includegraphics[scale=0.42]{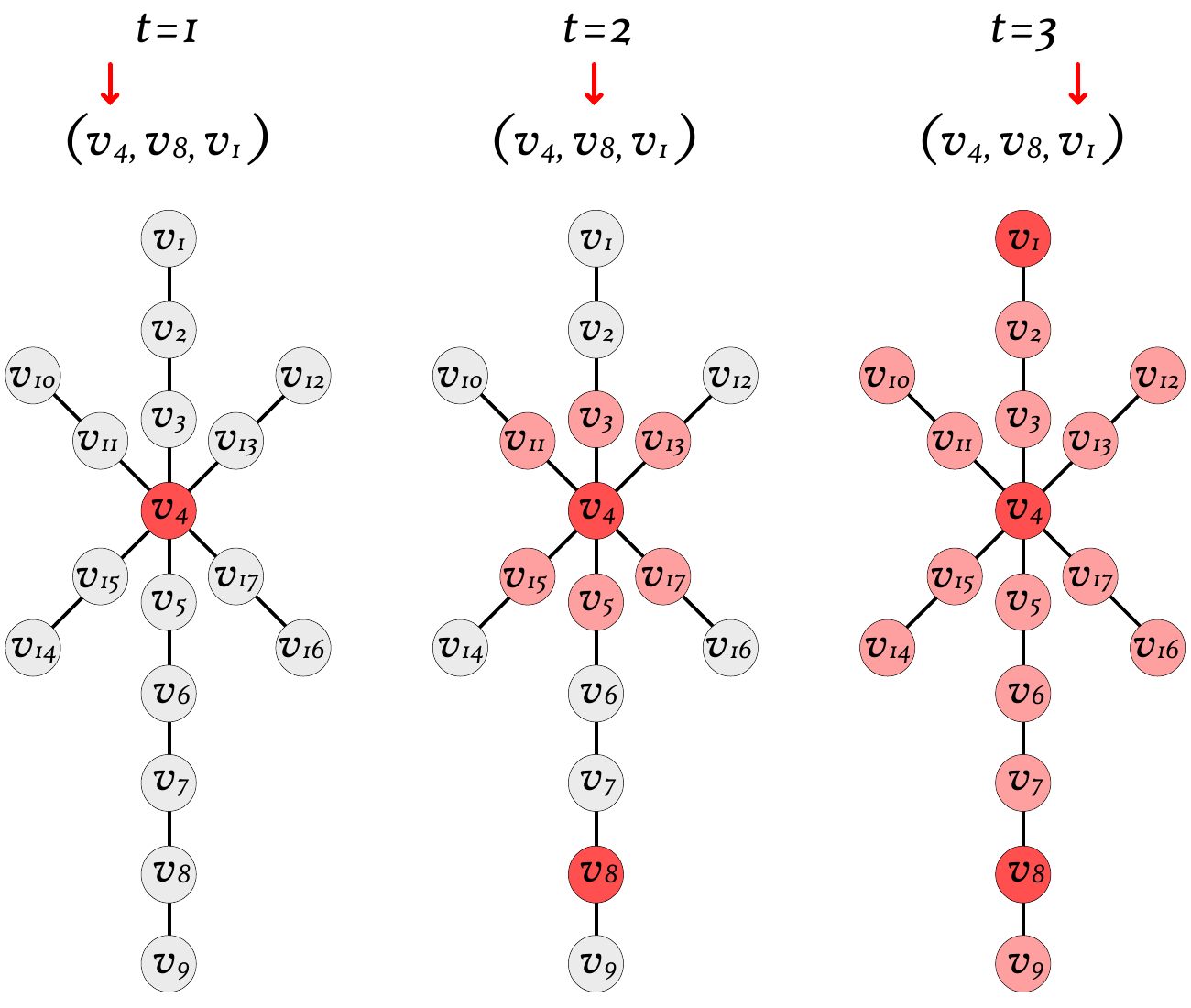}
	\caption{Burning process of the optimal burning sequence $(v_4,v_8,v_1)$; notice that $b(G)=3$. At each discrete time step, the \textit{fire} propagates to the neighbors of the previously burned vertices, and the next vertex in the sequence gets burned too. In the end, all vertices are burned.}
	\label{fig:1}
\end{figure}

The optimization version of the graph burning problem (GBP) is an NP-hard minimization problem that helps quantify a graph's vulnerability to contagion. It was introduced as a social contagion model by Bonato et al. \cite{bonato2014burning} and as a processor communication problem by Alon \cite{alon1992transmitting}. Nevertheless, it is an abstract problem with application in many other contexts, particularly those related to information or influence spreading \cite{gautam2022faster,chandrasekaran2022computational,rajeshwari2024revolutionizing}. In detail, GBP receives a graph $G$ as input and seeks a burning sequence of minimum length, where a burning sequence is a list of vertices that burns all the graph's vertices by the end of the burning process. This process consists of burning the neighbors of the previously burned vertices and the current vertex in the sequence at discrete time steps. Figure \ref{fig:1} shows the burning process of the optimal burning sequence of a given input graph. The length of the optimal burning sequence of a graph $G$ is known as its burning number and is denoted by $b(G)$.

The remaining part of the paper is organized as follows. Section \ref{sec2} summarizes the related work, introduces an integer linear program (ILP) for GBP, and lists the main definitions used in the following sections. Section \ref{sec3} shows how GBP reduces to a series of clustered maximum coverage problems (CMCP). Section \ref{sec4} introduces two more ILPs, a $(1/2)$-approximation algorithm for CMCP, and a deterministic greedy heuristic for GBP based on the approximation algorithm for CMCP. The proposed heuristic is conceptually simple and easy to implement. Besides, it performed unexpectedly well on benchmark and synthetic graphs, as reported in Section \ref{sec:exps}.
	
	\section{Related work}
	\label{sec2}
	
Many of the theoretical and practical aspects of GBP have been deeply explored in the last years. These include complexity analysis on specific graph classes, seeking tighter upper bounds for the burning number, mathematical modeling, and algorithm design. For a comprehensive survey, the reader can refer to the paper of Bonato \cite{bonato2020survey}.

GBP has an interesting relationship with other NP-hard problems. For instance, the minimum dominating set problem can be seen as the GBP with a parallel propagation process, where all vertices are burned in one discrete time step \cite{allan1978domination}. Besides, the known approximation algorithms for GBP are very similar to those for the vertex k-center problem \cite{garcia2019approximation,bonato2019approximation,garcia2022burning}, which reduces to a series of minimum dominating set problems and, by transitivity, to a series of minimum set cover problems \cite{minieka1970m,garey}. This paper shows how GBP is also related to the CMCP, a generalization of the NP-hard maximum coverage problem (MCP) \cite{hochbaum1998analysis,feige1998threshold}.

Regarding heuristics for the GBP, most of them rely on exploiting centrality measures, local search, or evolutionary computing. Therefore, they are relatively complicated and can be very time consuming. Among these are a centrality based genetic algorithm \cite{nazeri2023centrality} and different heuristics based on eigenvector centrality \cite{vsimon2019heuristics,gautam2022faster}. In constrast to these heuristics, the one introduced in this paper is simpler, is supported by deductive proofs, and has a good empirical performance, i.e., it found most of the optimal solutions of benchmark datasets in reasonable amounts of time.

Regarding mathematical models for the GBP, an integer linear program (ILP) and two constraint satisfaction problems (CSPs) have been reported \cite{garcia2022graph}. In this paper, we introduce three novel ILPs for GBP with a much better performance when implemented on a commercial optimization solver. Expressions (\ref{ILP1:1}) to (\ref{ILP1:6}) show the first of these ILPs, named ILP-PROP because it is based on the propagation process of the GBP. In this model, $G=(V,E)$ is the input graph, $n=|V|$, $m=|E|$, and $V=\{v_1,v_2,...,v_n\}$. It has $2Un+U$ binary decision variables and $2Un+U$ constraints, where $U$ is an upper bound on $b(G)$. Variables $b_{i,j}$ (resp., $s_{i,j}$) codify which vertices $v_i$ are burned (resp., added to the sequence) at each discrete time step $j$. Specifically, $b_{i,j}=1$ (resp., $s_{i,j}=1$) if vertex $v_i$ is burned (resp., is in the sequence) at the discrete time step $j$; $b_{i,j}=0$ (resp., $s_{i,j}=0$) otherwise. If all vertices are burned by step $j$, $x_j=0$; otherwise, $x_j=1$.


\begin{align}
	\textbf{min} & \ \ \sum\limits_{j\in [1,U]} x_{j}+1 &  \label{ILP1:1} \\
	\textbf{s.t.}
	& \ \ x_j \ge 1 - b_{i,j} & \forall v_i \in V, \forall j \in [1,U]  \label{ILP1:2}\\
	& \ \ b_{i,j} \le s_{i,j} + \sum \limits_{v_k\in N[v_i]} b_{k,j-1} &  \forall v_i \in V, \forall j \in [1,U]  \label{ILP1:3}\\		
	& \ \ \sum \limits_{v_i \in V} s_{i,j} = 1 &   \forall j \in [1,U]  \label{ILP1:4}\\
	\textbf{where}
	& \ \ s_{i,j} \ , \ b_{i,j} \ , \ x_j \in \{0,1\} &   \forall v_i \in	V, \forall j \in [1,U]  \label{ILP1:5} \\
	& \ \ b_{i,0} = 0 &  \forall v_i \in V  \label{ILP1:6}
\end{align}

Constraints (\ref{ILP1:2}) set $x_j=1$ if some vertex is not burned at the discrete time step $j$; since the Objective function (\ref{ILP1:1}) is minimized, $x_j$ is set to zero when all vertices are burned at the discrete time step $j$. This way, the objective function seeks to burn all vertices as soon as possible, i.e., at the smallest index $j$. Constraints (\ref{ILP1:3}) guarantee that a vertex cannot be burned unless it is into the sequence, was already burned, or has some burned neighbor. Finally, Constraints (\ref{ILP1:4}) indicate that 1 vertex must be added into the sequence at each discrete time step. The solution is codified into variables $s_{i,j}$, and $b(G)$ equals the smallest index $j$ at which all vertices are burned; namely, $b(G)$ is the value of the objective function; notice that the extra 1 in the objective function comes from the fact the smallest index $j$ at which all vertices get burned has $x_j=0$.

Before finishing this section, let us list the basic definitions used in the remaining part of the paper.\\


\begin{definition}
	\label{def:1}
	A graph $G=(V,E)$ is an ordered pair, where $V$ is the set of vertices and $E$ is the set of edges, a set of 2-element subsets of $V$.
\end{definition}

\vspace{0.15cm}

\begin{definition}
	\label{def:2}
	The distance between two vertices in a graph, $d(u,v)$, is the length of their shortest path.
\end{definition}

\vspace{0.15cm}

\begin{definition}
	\label{def:3}
	The open neighborhood $N(v)$ of a vertex $v$ is its set of neighbors in $G$.
\end{definition}

\vspace{0.15cm}

\begin{definition}
	\label{def:4}
	The closed neighborhood $N[v]$ of a vertex $v$ is $N(v) \cup  \{v\}$.
\end{definition}

\vspace{0.15cm}

\begin{definition}
	\label{def:5}
	The $r^{th}$ power $G^r$ of a given graph $G=(V,E)$ is the graph that results from adding an edge between each pair of vertices in $V$ at a distance at most $r$.
\end{definition}

\vspace{0.15cm}

\begin{definition}
	\label{def:6}
	The $r^{th}$ open neighborhood $N_{r}(v)$ of a vertex $v$ is its set of neighbors in $G^r$.
\end{definition}

\vspace{0.15cm}

\begin{definition}
	\label{def:7}
	The $r^{th}$ closed neighborhood $N_{r}[v]$ of a vertex $v$ is $N_{r}(v) \cup \{v\}$.
\end{definition}

\vspace{0.15cm}

\begin{definition}
	\label{def:8}
	A burning sequence of a graph $G=(V,E)$ is an ordered list $(u_1,u_2,...,u_p) \in V^p$ such that the distance from every $v \in V$ to some vertex $u_i$ is at most $p-i$. The length of the burning sequence is $p$.
\end{definition}

\vspace{0.15cm}

\begin{definition}
	\label{def:9}
	Given an input graph, the GBP seeks a burning sequence of minimum length.
\end{definition}
	
\section{Burning and covering}
	\label{sec3}
	
This section shows how GBP can be reduced to a series of CMCPs.

\vspace{0.15cm}

\begin{definition}
	The clustered maximum coverage problem (CMCP) receives $p$ clusters as input, each containing subsets of a universe of elements. Its goal is to select one subset from each cluster so that the cardinality of their union is maximized.
\end{definition}

\vspace{0.15cm}

\begin{proposition}
	\label{prop:1}
	Maximum coverage problem (MCP) is a particular case of CMCP.
\end{proposition}

\begin{proof}
	MCP receives a collection of subsets and a positive integer $p$ as input. Its goal is to find at most $p$ subsets such that the cardinality of their union is maximized \cite{hochbaum1998analysis}. Notice that an instance for MCP can be transformed into one for CMCP by creating $p$ identical copies of the input collection. Thus, solving the resulting CMCP implies solving MCP.
\end{proof}

Since MCP is NP-hard, CMCP is NP-hard too. The best approximation ratio achievable for the former cannot be greater than $(1-\frac{1}{e})$ unless $P=NP$ \cite{hochbaum1998analysis,hochbaum2020approximation,feige1998threshold}. Thus, by Proposition \ref{prop:1}, CMCP has the same inapproximability results.

\vspace{0.15cm}

\begin{proposition}
	\label{prop:red}
	GBP can be reduced to a series of CMCPs.
\end{proposition}

\begin{proof}
	Let $V$ be the universe of elements for CMCP, where $G=(V,E)$ is the input graph for GBP. The clusters for the corresponding CMCP are $\mathcal{C}_{r-1}=\{N_{r-1}[v]  :  v \in V\}, \forall r \in [1,p]$, where $p$ is a guess on $b(G)$; breadth-first search \cite{kozen1992depth,cherkassky1996shortest} can be used to compute the distance between all pairs of vertices efficiently. An optimal solution for CMCP consists of one subset from each cluster. An optimal solution for GBP is obtained by properly concatenating the corresponding vertices of each selected subset (See Definition \ref{def:8}). For instance, the clusters for a path $P_4=(\{v_1,v_2,v_3,v_4\},\{\{v_1,v_2\},\{v_2,v_3\},\{v_3,v_4\}\})$, with $p=b(P_4)=2$, are $\mathcal{C}_0=\{N_0[v_1],N_0[v_2],N_0[v_3],N_0[v_4]\}$ and $\mathcal{C}_1=\{N_1[v_1],N_1[v_2],N_1[v_3],N_1[v_4]\}$. An optimal solution for this instance is $\{N_0[v_4],N_1[v_2]\}$, which covers all the vertices, and its corresponding optimal burning sequence is $(v_2,v_4)$. Naturally, solving CMCP with $p=b(G)$ is necessary for the reduction to be useful. To find that value, we can solve CMCP within a binary search between 1 and $n$ (See Algorithm \ref{alg:trans}). If $p\ge b(G)$, an optimal solution for CMCP must cover all the vertices (line 8). If the optimal solution for CMCP does not cover all vertices, then $p<b(G)$ (line 10). This way, $p=b(G)$ at some point in the binary search, and the optimal burning sequence is found.
\end{proof}

\begin{algorithm}
	\caption{GBP as a series of CMCPs}\label{alg:trans}
	\begin{algorithmic}[1]
		\Require A graph $G=(V,E)$
		\Ensure An optimal burning sequence
		\State $l  \leftarrow 1$
		\State $h \leftarrow |V|$
		\State $D \leftarrow$ Compute the distance matrix
		\While{$l \le h$}
		\State $p \leftarrow \lfloor (l + h) / 2 \rfloor$
		\State $S \leftarrow$ \text{Solve CMCP}$(G,p,D)$
		\If{$S$ burns all vertices}
		\State $h \leftarrow p-1$
		\Else 
		\State $l \leftarrow p+1$
		\EndIf
		\EndWhile \\
		\Return the best found burning sequence
	\end{algorithmic}
\end{algorithm}

To further explain this reduction, let us introduce an ILP for GBP as a CMCP (See Expressions (\ref{IP:1}) to (\ref{IP:4})); we refer to this ILP as ILP-CMCP. The clusters are constructed based on the input graph $G=(V,E)$ and a given integer $p$; let us assume that $V=\{v_1,v_2,...,v_n\}$. This program has $pn+n$ binary decision variables and $p+n$ constraints. Decision binary variables $x_{i,j}$ are used to indicate which subset from each cluster is part of the solution (See Constraints (\ref{IP:2})). Decision binary variables $b_j$ indicate which vertices the selected subsets cover. Since the Objective function (\ref{IP:1}) maximizes the sum over all $b_j$, we add Constraints (\ref{IP:3}) to avoid a vertex $v_j$ being marked as covered (i.e., $b_j=1$) if it is not an element of some of the selected subsets; notice that $|\{N_{i-1}[v] : v \in V\}|=n$ for each cluster $i \in [1,p]$. Of course, a burning sequence of length $p$ exists only if $p\ge b(G)$. If so, the burning sequence is codified into variables $x_{i,j}$. Explicitly, the burning sequence is $(v_{g(p)},v_{g(p-1)},...,v_{g(1)})$, where $g(i)$ is the index $j$ where $x_{i,j}=1$. In case $p<b(G)$, a burning sequence of such length does not exist.

\begin{align}
	\textbf{max} & \ \ \sum\limits_{v_j \in V} b_j &  \label{IP:1} \\
	\textbf{s.t.} & \ \ \sum\limits_{v_j \in V} x_{i,j}=1 &   \forall i \in [1,p]  \label{IP:2}\\
	& \ \ b_{j} \le \sum\limits_{i\in [1,p]} \ \ \sum_{v_k \in N_{i-1}[v_j]} x_{i,k} &   \forall v_j \in V \label{IP:3}\\
	\textbf{where}
	& \ \ x_{i,j} \ , \ b_j \in \{0,1\} &  \forall i \in [1,p], \forall v_j \in V  \label{IP:4}
\end{align}

ILP-CMCP aims at covering as many vertices as possible and requires knowing $b(G)$ in advance to be equivalent to the GBP. As already mentioned, such an issue can be solved by using a binary search. However, ILP-CMCP can also be modified to remove such requirements. We refer to the resulting ILP as ILP-COV because it is based on the coverage definition of the GBP (See Expressions (\ref{IP2:1}) to (\ref{IP2:7})).

\begin{align}
	\textbf{min} & \ \ \sum\limits_{i \in [1,U]} \sum\limits_{v_j \in V} x_{i,j} &  \label{IP2:1} \\
	\textbf{s.t.}         & \ \ \sum_{v_j \in V}x_{i,j} \le \sum_{v_j \in V}x_{i-1,j} & \forall i \in [1,U] \label{IP2:2}\\
	& \ \ \sum\limits_{v_j \in V} x_{i,j}\le 1 &   \forall i \in [1,U]  \label{IP2:3}\\
	& \ \ b_{j} \le \sum\limits_{i\in [1,U]} \ \ \sum_{v_k \in N_{i-1}[v_j]} x_{i,k} &   \forall v_j \in V \label{IP2:4}\\
	& \ \ \sum_{v_j \in V} b_j = n & \label{IP2:5}\\
	\textbf{where}
	& \ \ x_{i,j} \ , \ b_j \in \{0,1\} &  \forall i \in [1,U], \forall v_j \in V  \label{IP2:6}\\
	& \ \ \sum_{v_j \in V}x_{0,j}=1 & \label{IP2:7}
\end{align}

ILP-COV has the same binary variables defined for ILP-CMCP. However, instead of a guess $p$, it requires an upper bound $U$ on $b(G)$. The Objective function (\ref{IP2:1}) minimizes the number of vertices in the burning sequence. Thanks to Constraints (\ref{IP2:2}), (\ref{IP2:3}), and (\ref{IP2:7}), only 1 vertex for each valid coverage in $\{0,1,\dots,b(G)-1\}$ can be selected. By Constraints (\ref{IP2:4}), a vertex cannot be marked as covered if it is not in the $i-1$ neighborhood of some vertex with coverage capacity $i-1$. Finally, Constraint (\ref{IP2:5}) guarantees that all vertices are burned. ILP-COV has $Un+n$ binary variables and $2U+n+1$ constraints.

The next section introduces a greedy $(1/2)$-approximation algorithm for CMCP and its heuristic adaptation to GBP. This heuristic was tested in Section \ref{sec:exps}, and the solutions it returned were used as upper bound for ILP-COV, which was implemented in a commercial optimization solver to find optimal solutions. This way, we could evaluate the quality of the solutions returned by the proposed greedy algorithm.
	
\section{Greedy algorithms}
	\label{sec4}
	
This section is divided into two parts. The first introduces a greedy $(1/2)$-approximation algorithm for CMCP. This is used as a heuristic for GBP in the second part.

\subsection{A greedy $(1/2)$-approximation algorithm for CMCP}

\begin{figure}[h]
	\centering
	\includegraphics[scale=0.4]{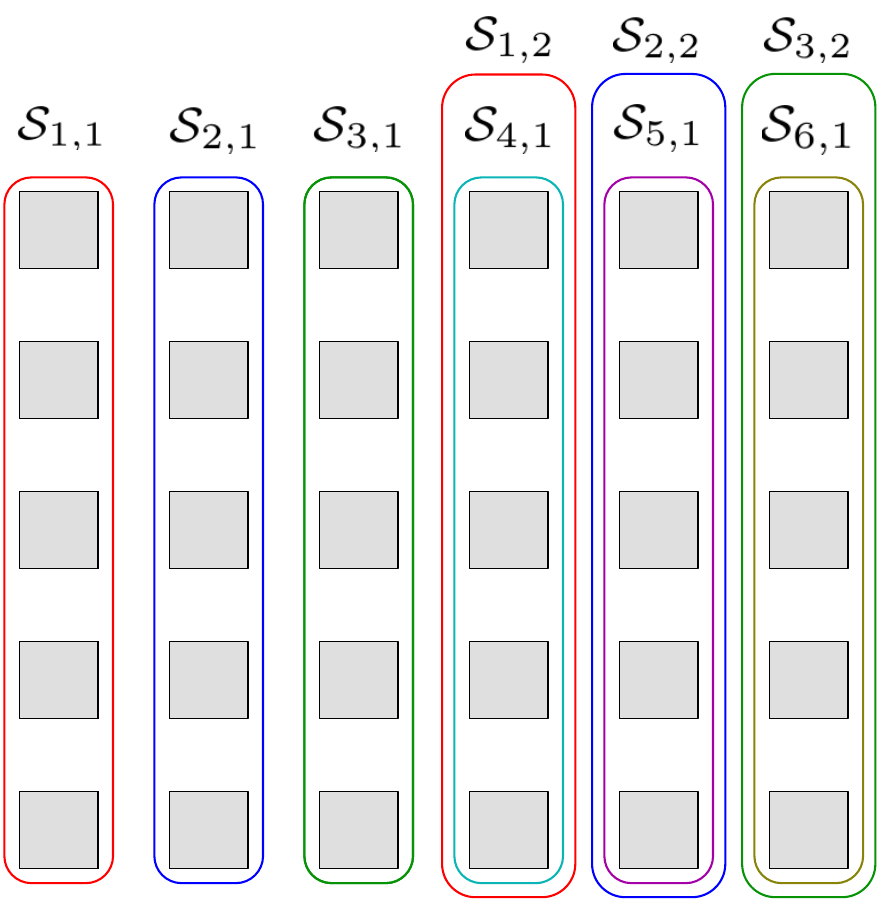}
	\caption{In this instance for CMCP, $p=6$, $\mathcal{C}_1 = \{\mathcal{S}_{1,1},\mathcal{S}_{1,2}\}$, $\mathcal{C}_2 = \{\mathcal{S}_{2,1},\mathcal{S}_{2,2}\}$, $\mathcal{C}_3 = \{\mathcal{S}_{3,1},\mathcal{S}_{3,2}\}$, $\mathcal{C}_4 = \{\mathcal{S}_{4,1}\}$, $\mathcal{C}_5 = \{\mathcal{S}_{5,1}\}$, and $\mathcal{C}_6 = \{\mathcal{S}_{6,1}\}$. The optimal solution $\{\mathcal{S}_{1,1}, \mathcal{S}_{2,1}, \mathcal{S}_{3,1}, \mathcal{S}_{4,1}, \mathcal{S}_{5,1}, \mathcal{S}_{6,1}\}$ covers $OPT=30$ squares. If the greedy approximation algorithm selects $\mathcal{S}_{1,2}$ at the first iteration, all the squares inside $\mathcal{S}_{1,1}$ get blocked because no other subsets from the remaining clusters cover them. If the algorithm selects $\mathcal{S}_{2,2}$ in the second iteration, all the squares inside $\mathcal{S}_{2,1}$ get blocked. Finally, if the algorithm selects $\mathcal{S}_{3,2}$ at the third iteration, all the squares inside $\mathcal{S}_{3,1}$ get blocked. Thus, the next three selected subsets won't be able to cover any uncovered squares and the returned solution will be $\{\mathcal{S}_{1,2},\mathcal{S}_{2,2},\mathcal{S}_{3,2},\mathcal{S}_{4,1},\mathcal{S}_{5,1},\mathcal{S}_{6,1}\}$, which covers $OPT/2=15$ squares.}
	\label{fig:3}
\end{figure}

The classical greedy approximation algorithm for MCP iteratively selects the subset that covers as many uncovered elements as possible. This algorithm achieves the best possible ratio of $(1-\frac{1}{e})$ \cite{hochbaum1998analysis,hochbaum2020approximation,feige1998threshold}. The greedy approximation algorithm for CMCP follows the same idea. Given $p$ clusters of subsets of a universe of elements, iteratively selects the subset that covers as many uncovered elements as possible among all the available clusters (See Algorithm \ref{alg:grc}). Every time a bad decision is made, some important elements might be blocked; namely, they won't be selected in future iterations. So, there will be some \textit{covered} and some \textit{blocked} elements at every iteration. Figure \ref{fig:3} exemplifies this situation, where only $OPT/2$ elements are covered by the returned feasible solution. In fact, $1/2$ is the approximation ratio of this algorithm (See Theorem \ref{tem:1}, which relies on Lemmas \ref{lem:1}, \ref{lem:2}, and \ref{lem:3}); Figure \ref{fig:3} serves as a tight example.

\begin{algorithm}
	\caption{Greedy $(1/2)$-approximation algorithm for CMCP}\label{alg:grc}
	\begin{algorithmic}[1]
		\Require $p$ clusters $\mathcal{C}_1,\mathcal{C}_2,...,\mathcal{C}_p$
		\Ensure A set $S$ containing one subset from each cluster
		\State $S, C \leftarrow \emptyset$
		\State $K \leftarrow \{1,2,...,p\}$
		\While{$K \not = \emptyset$}
		\State $k \leftarrow \argmax_{j \in K} \max_{\mathcal{S} \in \mathcal{C}_j} |\mathcal{S}-C|$
		\State $K \leftarrow K - \{k\}$  
		\State $A \leftarrow \argmax_{\mathcal{S} \in \mathcal{C}_k} |\mathcal{S}-C|$		
		\State $S \leftarrow S \cup \{A\}$
		\State $C \leftarrow C \cup A$
		\EndWhile \\
		\Return $S$
	\end{algorithmic}
\end{algorithm}

\vspace{0.15cm}

\begin{lemma}
	\label{lem:1}
	At each iteration of Algorithm \ref{alg:grc}, the number of newly blocked elements cannot exceed the number of newly covered elements. By ``newly'' we mean that those elements were not blocked or covered in previous iterations.
\end{lemma}

\begin{proof}
	Let $\mathcal{C}_1,\mathcal{C}_2,...,\mathcal{C}_p$ be the clusters of the input instance. Let $S^*=\{ \mathcal{S}_1^*,\mathcal{S}_2^*,...,\mathcal{S}_p^* \}$ be an optimal solution and $S'=\{ \mathcal{S}_1',\mathcal{S}_2',...,\mathcal{S}_p' \}$ be the feasible solution returned by Algorithm \ref{alg:grc}, where $\mathcal{S}_i^*,\mathcal{S}_i' \in \mathcal{C}_i$. A set $\mathcal{S}_j'$ with maximum coverage $|\mathcal{S}_j'-C|$ is selected at every iteration, where $C$ is the set of covered elements at previous iterations. If $\mathcal{S}_j' \not \in S^*$ (or equivalently, if $\mathcal{S}_j'\not =\mathcal{S}_j^*$), then in the worst scenario all the elements in $\mathcal{S}_j^*-C$ are not in the subsets from the remaining clusters; therefore, they get blocked. Since $|\mathcal{S}_j'-C|$ is maximal, $|\mathcal{S}_j^*-C| \le |\mathcal{S}_j'-C|$.
\end{proof}

Lemma \ref{lem:2} shows how the size of the optimal solution for a given instance $\Pi_\alpha$ is related to the size of the optimal solution of smaller instances $\Pi_\beta$ that result from removing some cluster and some specific elements from $\Pi_\alpha$. This lemma will be useful to prove the next propositions.

\begin{lemma}
	\label{lem:2}
	Let $\Pi_\alpha$ be an arbitrary instance for the GBP with clusters $\mathcal{C}_1, \mathcal{C}_2,...,\mathcal{C}_p$, an optimal solution size of $OPT(\Pi_\alpha)$, and an optimal solution $S^*=\{ \mathcal{S}_1^*,\mathcal{S}_2^*,...,\mathcal{S}_p^* \}$, where $\mathcal{S}_1^* \in \mathcal{C}_1$, $\mathcal{S}_2^* \in \mathcal{C}_2$, and so on. Let $\mathcal{S}_a$ be an arbitrary subset from an arbitrary cluster $\mathcal{C}_a$ and let $\mathcal{S}_a^*$ be the one element in $\mathcal{C}_a \cap S^*$. Besides, let $\Pi_\beta$ be an instance constructed as follows: it has the same clusters as $\Pi_\alpha$, except $\mathcal{C}_a$, and $\mathcal{S}_a \cap \cup_{\{\mathcal{S}\in \mathcal{C}_j : j \in [1,p]-\{a\} \wedge \mathcal{S}\in S^* \}} \mathcal{S}$ is removed from the universe of elements. The optimal solution size of $\Pi_\beta$ is $OPT(\Pi_\alpha)-y-c$, where $y$ is the number of elements only $\mathcal{S}_a^*$ can cover (these would be the blocked elements in Algorithm \ref{alg:grc}), i.e., 
	
	\begin{equation}
		y=|\mathcal{S}_a^* - \mathcal{S}_a^*\cap \cup_{\{\mathcal{S}\in \mathcal{C}_j : j \in [1,p]-\{a\} \wedge \mathcal{S}\in S^* \}} \mathcal{S}| \ \ ,
		\label{eq:lem2:1}
	\end{equation}
	
	and $c$ is the number of elements in $\mathcal{S}_a$ that would contribute to $OPT(\Pi_\beta)$ if they were not removed, i.e.,
	
	\begin{equation}
		c=|\mathcal{S}_a \cap \cup_{\{\mathcal{S}\in \mathcal{C}_j : j \in [1,p]-\{a\} \wedge \mathcal{S}\in S^* \}} \mathcal{S}| \ \ .
		\label{eq:lem2:2}
	\end{equation}
	
\end{lemma}

\begin{proof}
	Let $\Pi_\beta^\prime$ be the instance that results by only removing $\mathcal{C}_a$ from $\Pi_\alpha$.
	Notice that $OPT(\Pi_\beta^\prime)$ cannot be smaller than $OPT(\Pi_\alpha)-y$ because this is the number of elements covered by all the subsets in $S^*- \{\mathcal{S}_a^*\}$. Thus, $OPT(\Pi_\beta^\prime) \ge OPT(\Pi_\alpha)-y$. In case $OPT(\Pi_\beta^\prime) > OPT(\Pi_\alpha)-y$, then $OPT(\Pi_\alpha)<OPT(\Pi_\beta^\prime)+y$, which contradicts the optimality of $OPT(\Pi_\alpha)$. Therefore, $OPT(\Pi_\beta^\prime) = OPT(\Pi_\alpha)-y$. By a similar argument, $OPT(\Pi_\beta)$ cannot be different from $OPT(\Pi_\beta^\prime)-c$. Therefore, $OPT(\Pi_\beta)=OPT(\Pi_\beta^\prime)-c=OPT(\Pi_\alpha)-y-c$.
\end{proof}

\begin{figure}[h]
	\centering
	\includegraphics[scale=0.4]{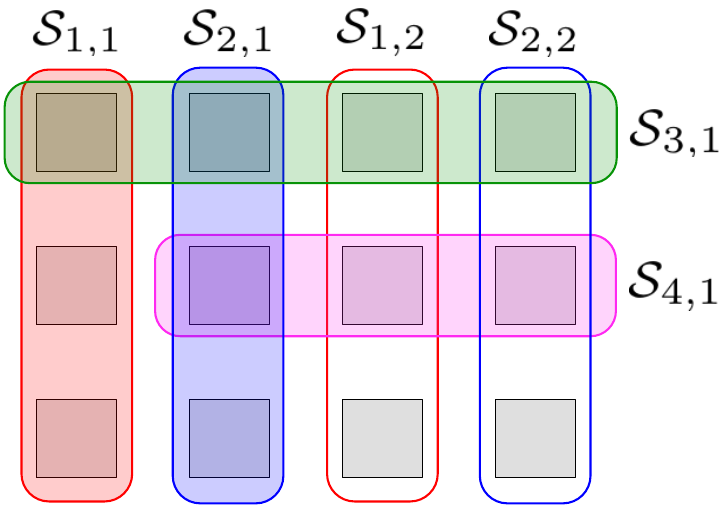}
	\caption{An instance $\Pi_\alpha$ with $p=4$ and $OPT(\Pi_\alpha)=10$.}
	\label{fig:ex1}
\end{figure}

To better understand Lemma \ref{lem:2}, consider an instance $\Pi_\alpha$ with $p=4$, $\mathcal{C}_1=\{\mathcal{S}_{1,1},\mathcal{S}_{1,2}\}$, $\mathcal{C}_2=\{\mathcal{S}_{2,1},\mathcal{S}_{2,2}\}$, $\mathcal{C}_3=\{\mathcal{S}_{3,1}\}$, and  $\mathcal{C}_4=\{\mathcal{S}_{4,1}\}$; it has an optimal solution $\{\mathcal{S}_{1,1},\mathcal{S}_{2,1},\mathcal{S}_{3,1},\mathcal{S}_{4,1}\}$ and $OPT(\Pi_\alpha)=10$ (See Figure \ref{fig:ex1}).

\begin{figure}[h]
	\centering
	\includegraphics[scale=0.4]{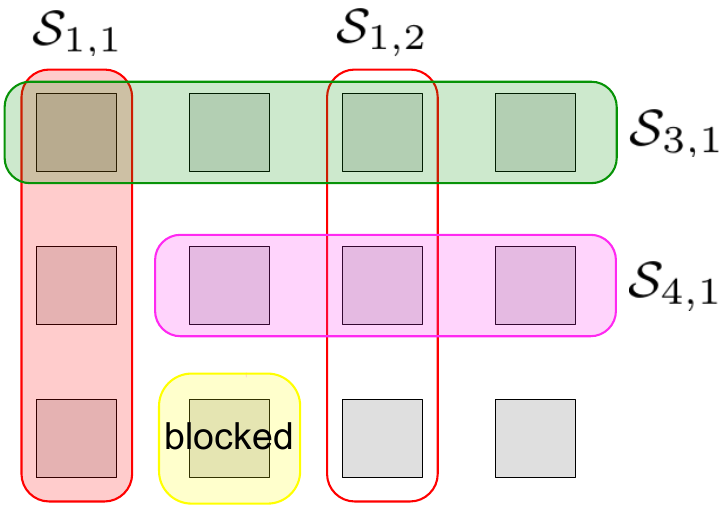}
	\caption{An instance $\Pi_\beta^\prime$ with $p=3$ and $OPT(\Pi_\beta^\prime)=9$. $\Pi_\beta^\prime$ results from removing $\mathcal{C}_2$ from $\Pi_\alpha$.}
	\label{fig:ex2}
\end{figure}

By selecting $\mathcal{S}_{2,2}$ as the ``arbitrary'' subset from the ``arbitrary'' cluster $\mathcal{C}_2$, removing $\mathcal{C}_2$ from $\Pi_\alpha$ results in $\Pi_\beta^\prime$, which has $p=3$, $\mathcal{C}_1=\{\mathcal{S}_{1,1},\mathcal{S}_{1,2}\}$, $\mathcal{C}_3=\{\mathcal{S}_{3,1}\}$, and  $\mathcal{C}_4=\{\mathcal{S}_{4,1}\}$ (See Figure \ref{fig:ex2}); it has an optimal solution $\{\mathcal{S}_{1,1},\mathcal{S}_{3,1},\mathcal{S}_{4,1}\}$ and $OPT(\Pi_\beta^\prime)=OPT(\Pi_\alpha)-y=10-1=9$, where $y=1$ is the number of blocked elements, i.e., elements in $\mathcal{S}_{2,1} - \mathcal{S}_{2,1} \cap (\mathcal{S}_{1,1} \cup \mathcal{S}_{3,1} \cup \mathcal{S}_{4,1})$ (See Eq. (\ref{eq:lem2:1})).

\begin{figure}[h]
	\centering
	\includegraphics[scale=0.4]{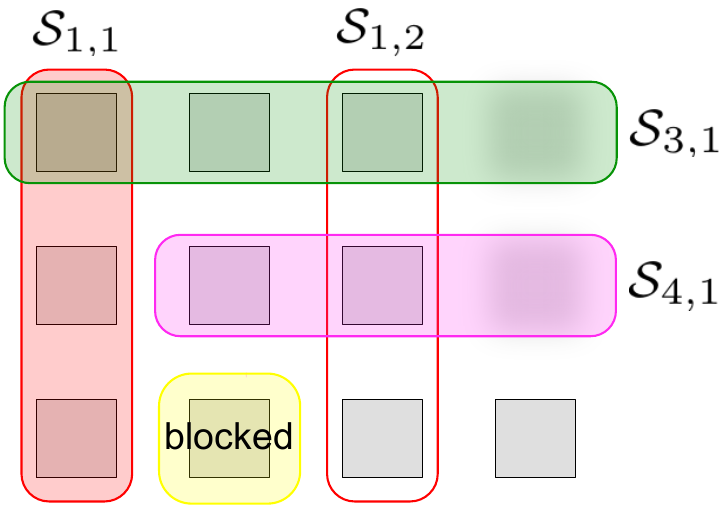}
	\caption{An instance $\Pi_\beta$ with $p=3$ and $OPT(\Pi_\beta)=7$. $\Pi_\beta$ results from removing some specific elements from $\Pi_\beta^\prime$.}
	\label{fig:ex3}
\end{figure}

Afterward, by removing the elements in $S_{2,2}$ that contribute to $OPT(\Pi_\beta^\prime)$ from $\Pi_\beta^\prime$, we get $\Pi_\beta$ (See Figure \ref{fig:ex3}). Thus, $OPT(\Pi_\beta)=OPT(\Pi_\beta^\prime)-c=OPT(\Pi_\alpha)-y-c=10-1-2=7$, where $c$ is the cardinality of $\mathcal{S}_{2,2} \cap (\mathcal{S}_{1,1} \cup \mathcal{S}_{3,1} \cup \mathcal{S}_{4,1})$ (See Eq. (\ref{eq:lem2:2})).

Algorithm \ref{alg:grc} iteratively prunes the input instance by removing a specific cluster; hence the relevance of Lemma \ref{lem:2}. In more detail, by the end of every iteration $i$ of Algorithm \ref{alg:grc}, an instance $\Pi_i$ with optimal solution size $OPT(\Pi_i)$ is obtained, being $\Pi_0$ the original input instance. Each $\Pi_i$ results from pruning $\Pi_{i-1}$ as follows: remove the cluster $\mathcal{C}_{l_i}$ that contains the largest subset $\mathcal{S}_{l_i}$ among all available clusters (line 5) and remove all elements in $\mathcal{S}_{l_i}$ from the universe of elements (lines 4, 6, and 8 of Algorithm \ref{alg:grc}). Using this notation, a lower bound on $OPT(\Pi_i)$ is derived in the next lemma.

\begin{lemma}
	\label{lem:3}
	$z_i=z_{i-1}-y_i-x_i$ is a lower bound for $OPT(\Pi_i)$, where $y_i$ is the number of newly blocked elements at iteration $i$, $x_i$ is the number of newly covered elements at iteration $i$, and $z_0=OPT(\Pi_0)$.
\end{lemma}

\begin{proof}
	By definition, 
	\begin{equation}
		z_i=z_{i-1}-y_i-x_i \ \ .	
		\label{eq:lem3:1}
	\end{equation}
	By Lemma \ref{lem:2}, 
	\begin{equation}
		OPT(\Pi_i) = OPT(\Pi_{i-1})-y_i-c_i	\ \ ,
		\label{eq:lem3:2}
	\end{equation}
	where $y_i$ is the number of newly blocked elements (See Eq. (\ref{eq:lem2:1})) and $c_i$ is the cardinality of a subset of $\mathcal{S}_{l_i}-C$ (See Eq. (\ref{eq:lem2:2})). Therefore, 
	
	\begin{equation}
		c_i\le |\mathcal{S}_{l_i}-C| = x_i	\ \ ,
		\label{eq:lem3:3}
	\end{equation}
	
	where $C$ is the set of covered elements at previous iterations. Notice that keeping track of $C$ at lines 4, 6, and 8 of Algorithm \ref{alg:grc} is equivalent to removing all the previously covered elements from each $\Pi_i$. Since $z_0=OPT(\Pi_0)$, Eq. (\ref{eq:lem3:2}) leads to
	
	\begin{equation}
		OPT(\Pi_1) = z_0-y_1-c_1 \ \ .
		\label{eq:lem3:4}
	\end{equation}
	
	By Eq. (\ref{eq:lem3:3}), $c_1\le x_1$, which along with Eq. (\ref{eq:lem3:1}) and Eq. (\ref{eq:lem3:4}) leads to
	
	\begin{equation}
		z_1=z_0-y_1-x_1\le z_0-y_1-c_1=OPT(\Pi_1) \ \ .
		\label{eq:lem3:5}
	\end{equation}
	
	Therefore,
	
	\begin{equation}
		z_1\le OPT(\Pi_1) \ \ .
		\label{eq:lem3:6}
	\end{equation}
	
	For $i=2$, Eq. (\ref{eq:lem3:1}) and Eq. (\ref{eq:lem3:6}) lead to
	
	\begin{equation}
		z_2=z_1-y_2-x_2\le OPT(\Pi_1)-y_2-x_2 \ \ .
		\label{eq:lem3:7}
	\end{equation}
	
	Then, Eq. (\ref{eq:lem3:7}), Eq. (\ref{eq:lem3:2}) and $c_2-x_2\le 0$ (Eq. (\ref{eq:lem3:3})) lead to
	
	\begin{equation}
		z_2\le OPT(\Pi_2)+c_2-x_2 \le OPT(\Pi_2)\ \ .
		\label{eq:lem3:8}
	\end{equation}

	Thus,

	\begin{equation}
		z_2\le OPT(\Pi_2)\ \ .
		\label{eq:lem3:8:b}
	\end{equation}
	
	For $i \in [3,p]$, the pattern continues:
	
	\begin{equation}
		z_i = z_{i-1} - y_i -x_i \le OPT(\Pi_{i-1})- y_i -x_i \ \ ,
		\label{eq:lem3:9}
	\end{equation}
	
	\begin{equation}
		z_i \le OPT(\Pi_i) + c_i -x_i \le OPT(\Pi_i) \ \ ,
		\label{eq:lem3:10}		
	\end{equation}
	
	\begin{equation}
		z_i \le OPT(\Pi_i) \ \ .
		\label{eq:lem3:11} 
	\end{equation}
	
\end{proof}

\begin{theorem}
	\label{tem:1}
	The greedy approximation algorithm for CMCP (Algorithm \ref{alg:grc}) has an approximation ratio of $1/2$.
\end{theorem}

\begin{proof}
	Let $\Pi_0$ be an input instance for the GBP. Let $OPT(\Pi_0)$ be the size of the optimal solution, let $x_i$ be the number of newly covered elements at iteration $i$, and let $y_i$ be the number of newly blocked elements at iteration $i$. Since $x_1$ is maximal,
	
	\begin{equation}
		\label{eq:c0}
		x_1 \ge \frac{OPT(\Pi_0)}{p} \ \ \ \ ;
	\end{equation}	
	
	otherwise, $px_1<OPT(\Pi_0)$, which means that $p$ disjoint largest subsets would not be able to cover $OPT(\Pi_0)$ elements, which is a contradiction. Using Lemma \ref{lem:3}, this argument can be generalized to all iterations $i \in [1,p]$ in terms of $z_i$,
	
	\begin{equation}
		\label{eq:c1}
		x_i \ge \frac{OPT(\Pi_{i-1})}{p-(i-1)} \ge \frac{z_{i-1}}{p-(i-1)} \ \ \ \ ,
	\end{equation}
	
	Notice that $z_0=OPT(\Pi_0)$ and $z_p=0$ because all $z_{p-1}$ vertices are covered at the last iteration. Besides, by Lemma \ref{lem:1}, no more than $x_i$ elements can be blocked at each iteration:
	
	\begin{equation}
		\label{eq:c3}		
		y_i \le x_i \ \ \ \ .
	\end{equation}
	
	With these observations, the theorem is proved by contradiction. Let us assume that the algorithm returns a solution that covers less than $OPT(\Pi_0)/2$ elements, i.e.,
	
	\begin{equation}
		\label{eq:c4}
		\sum_{i=1}^{p}x_i < \frac{OPT(\Pi_0)}{2} \ \ \ \ .
	\end{equation}
	
	Eq. (\ref{eq:c3}) and Eq. (\ref{eq:c4}) lead to
	
	\begin{equation}
		\label{eq:c5}
		\sum_{i=1}^{p}y_i \le \sum_{i=1}^{p}x_i < \frac{OPT(\Pi_0)}{2} \ \ \ \ .
	\end{equation}
	
	Then, Eq. (\ref{eq:lem3:1}) and Eq. (\ref{eq:c5}) lead to
	
	\begin{equation}
		\label{eq:c6}	
		\sum_{i=1}^{p}z_i >  \sum_{i=1}^{p}z_{i-1} - OPT(\Pi_0)\ \ \ \ .
	\end{equation}
	
	Notice that $\sum_{i=1}^{p}z_{i-1} - \sum_{i=1}^{p}z_{i}=z_0-z_p $. Since $z_0 = OPT(\Pi_0)$ and $z_p=0$, Eq. (\ref{eq:c6}) reduces to
	
	\begin{equation}
		\label{eq:c7}	
		OPT(\Pi_0) = z_0 - z_p < OPT(\Pi_0) \ \ \ \ ,
	\end{equation}
	
	which is a contradiction. Therefore, $\sum_{i=1}^{p}x_i \ge OPT(\Pi_0)/2$. Namely, the greedy approximation algorithm for CMCP has an approximation ratio of $1/2$.
	
\end{proof}

In the next section, we adapt Algorithm \ref{alg:grc} to GBP, resulting in a heuristic algorithm.
	
\subsection{A greedy heuristic for GBP}
	
Algorithm \ref{alg:gr} shows the proposed deterministic greedy heuristic (Gr) for the GBP, which implements the greedy approximation algorithm for CMCP. In a nutshell, starting from the clusters with larger subsets, i.e., from $p-1$ to 0 (line 4), Gr iteratively selects the corresponding vertex of the subset that covers as many uncovered elements as possible (lines 8 and 9); notice that `$+$' adds a vertex at the right side of the sequence, which is initially empty. In more detail, Gr receives as input a graph $G=(V,E)$, a guess $p$ on $b(G)$, and the distance matrix $D$, which can be computed independently using breadth-first search. Then, Gr computes the closed neighborhood $N_{p-1}[v]$ of each $v \in V$. In other words, it computes the adjacency list of $G^{p-1}$ (line 3). Afterward, it iteratively selects $p$ vertices that correspond to the subsets that cover as many uncovered vertices as possible in graphs $G^r$, where $r$ goes from $p-1$ to $0$ (lines 4 to 11). Notice that the adjacency list is updated at each iteration, containing uncovered vertices at a distance at most $r$ from each $v \in V$ (line 6). Besides, every subset $N_i[v]$ of each cluster is a superset of each $N_j[v]$, where $j<i$. Thus, the corresponding vertex of a subset with maximum coverage among all the clusters is selected at each iteration. Therefore, this is a correct implementation of Algorithm \ref{alg:grc}. Since the distance matrix $D$ is known, updating the adjacency list can be done in at most $\mathcal{O}(n^2)$ steps at each iteration. Thus, the overall complexity of Gr is $\mathcal{O}(pn^2)$; however, if we consider the time required to compute all-pairs' shortest path, its complexity is $\mathcal{O}(mn+pn^2)$ as stated in the abstract. By Theorem \ref{tem:1}, Proposition \ref{prop:4} follows.

\begin{algorithm}
	\caption{Greedy burning (Gr)}\label{alg:gr}
	\begin{algorithmic}[1]
		\Require A graph $G=(V,E)$, its distance matrix $D$, and a guess $p$ on $b(G)$
		\Ensure A sequence $S$
		\State $S \leftarrow ()$
		\State $B \leftarrow \emptyset$
		\State $N_{p-1} \leftarrow$ Adjacency list of $G^{p-1}$			\For{$r=p-1$ \textbf{to} $0$}
		\If{$r<p-1$}
		\State $N_r \leftarrow$ Update the adjacency list for $G^{r}$
		\EndIf
		\State $v \leftarrow \argmax_{u \in V} |N_r[u]-B|$
		\State $S \leftarrow S + v$
		\State $B \leftarrow B \cup  N_r[v]$
		\EndFor \\
		\Return $S$
	\end{algorithmic}
\end{algorithm}

\vspace{0.15cm}

\begin{proposition}
	\label{prop:4}
	If the subsets Gr selects cover less than $n/2$ vertices, then $p<b(G)$.
\end{proposition}

\begin{proof}
	By Theorem \ref{tem:1}, Algorithm \ref{alg:grc} returns a solution that covers at least $OPT/2$ elements, where $OPT$ is the optimal solution size. By definition, a burning sequence burns $n$ vertices. So, if $p \ge b(G)$, $OPT=n$, and since Gr correctly implements Algorithm \ref{alg:grc}, it must return a solution that covers at least $n/2$ vertices. By contrapositive, if Gr returns a solution that covers less than $n/2$ vertices, then $p<b(G)$.
\end{proof}

For Gr to be useful, we must know $b(G)$. We can try different guesses $p$ to tackle the issue of not knowing this value in advance. The simpler way is by trying all $p \in [1,n]$. However, a binary search is more efficient. Although Proposition \ref{prop:4} helps find a lower bound for $b(G)$, we didn't find it useful in practice. Therefore we didn't integrate it into the binary search (See Algorithm \ref{alg:bs}), which we now explain. First, the distance between every pair of vertices is computed in $\mathcal{O}(mn+n^2)$ steps using breadth-first search (line 1). Then, the lower and upper bounds are computed using the burning farthest-first (BFF) approximation algorithm for GBP \cite{garcia2022burning}, which runs in  $\mathcal{O}(n^2)$ steps if the computation of all-pairs' shortest path is not considered; Algorithm \ref{alg:bff} shows a detailed pseudocode for BFF using a first-in-first-out (FIFO) queue. Since BFF returns a solution $S_0$ (line 2) of length at most $3b(G)-2$ \cite{garcia2022burning}, the lower bound on $b(G)$ can be safely set to $\lceil (s(S_0)+2)/3 \rceil$ (line 3) and the upper bound to $s(S_0)-1$ (line 4), where $s(S_0)$ is the size of $S_0$. Then, Gr is executed with parameters $G$, $p$, and $D$ (lines 6 and 7); GrP will be introduced later. Afterward, every time Gr returns a burning sequence, $h$ is set to $p-1$; otherwise, $l$ is set to $p+1$. This way, Gr is executed $\mathcal{O}(\log n)$ times with different guesses on $p$ that might approach $b(G)$. Using Gr at line 7, and since $p\le n$, the complexity of Algorithm \ref{alg:bs} is $\mathcal{O}(n^3 \log n)$.

\begin{algorithm}
	\caption{Binary search for GBP}\label{alg:bs}
	\begin{algorithmic}[1]
		\Require A graph $G=(V,E)$
		\Ensure A burning sequence
		\State $D \leftarrow$ Compute the distance matrix		
		\State $S_0 \leftarrow \text{BFF}(G)$
		\State $l  \leftarrow \lceil(s(S_0)+2)/3\rceil$
		\State $h \leftarrow s(S_0)-1$
		\While{$l \le h$}
		\State $p \leftarrow \lfloor (l + h) / 2 \rfloor$
		\State $S \leftarrow$ Gr$(G,p,D)$ or GrP$(G,p,D)$
		\If{$S$ burns all vertices}
		\State $h \leftarrow p-1$
		\Else 
		\State $l \leftarrow p+1$
		\EndIf
		\EndWhile \\
		\Return the best found burning sequence
	\end{algorithmic}
\end{algorithm}

\begin{algorithm}
	\caption{Burning Farthest-First (BFF)}\label{alg:bff}
	\begin{algorithmic}[1]
		\Require A graph $G=(V,E)$ and its distance matrix $D$
		\Ensure A burning sequence $S$ of length at most $3b(G)-2$
		\State $q \leftarrow $ Empty FIFO queue
		\State $v \leftarrow $ Any vertex from $V$
		\State $S \leftarrow (v)$
		\State $q.push(v)$
		\State $B \leftarrow \{v\}$
		\For{$u \in V$}
		\State $dist(u) \leftarrow d(u,v)$
		\EndFor
		\While{$B \not= V$}
		\State $q\_size \leftarrow q.size()$
		\For{$j=1$ to $q\_size$}
		\State $v \leftarrow q.pop()$
		\For{$u \in N(v) - B $}
		\State $q.push(u)$
		\State $B \leftarrow B \cup \{u\}$
		\EndFor
		\EndFor
		\State $v \leftarrow \argmax_{u \in V} dist(u)$
		\State $S \leftarrow S + v$
		\State $q.push(v)$
		\State $B \leftarrow B \cup \{v\}$
		\For{$u \in V$}
		\If{$d(u,v)<dist(u)$}
		\State $dist(u) \leftarrow d(u,v)$
		\EndIf
		\EndFor
		\EndWhile \\
		\Return $S$
	\end{algorithmic}
\end{algorithm}

Using BFF at line 2 of Algorithm \ref{alg:bs} guarantees that a burning sequence of length at most $3b(G)-2$ is found \cite{garcia2022burning}. Nevertheless, in the worst scenario, Gr might not find a better solution than that given by BFF. In fact, it might not return a burning sequence for paths and cycles (See Proposition \ref{prop:3} and Corollary \ref{cor:1}). Since there are exact $\mathcal{O}(n)$ algorithms for GBP over these graphs' families \cite{bonato2016burn}, the behavior of Gr over them might hint at its theoretical limitations.

\vspace{0.15cm}

\begin{proposition}
	\label{prop:3}
	Gr might not return a burning sequence for a path $P$.
\end{proposition}

\begin{proof}
	It is well known that $b(P_n)=\lceil n^{1/2} \rceil$ for any path $P_n$ of order $n$ \cite{bonato2014burning}. So, given a path $P_n$, we can run Gr with $p=b(P_n)$, which is equivalent to execute Algorithm \ref{alg:bs} with $l=h=b(P_n)$. For the rest of the proof, let us consider paths $P_n$ with the largest possible order for any given burning number, namely, paths $P_n$ with $n=z^2$ for any $z \in \mathbb{Z}^+$. Also, observe that the maximum number of vertices a vertex $v$ can cover in a path is $|N_{r}[v]|=2r+1$, for any given $r$. 
	
	The proof is divided into three parts. The first considers the initial iteration of the \textit{for} loop of Gr ($r=p-1$); the second considers iterations $r=p-2$ to $r=2$, and the last one considers iterations $r=1$ and $r=0$. In the first iteration, $r=p-1$, Gr selects a vertex that covers exactly $2r+1$ vertices because the path's length, $(r+1)^2$, is bounded by $2r+1$, i.e.,
	
	\begin{equation}
		\label{eq:1}
		|P_n| = (r+1)^2 \ge 2r+1 \ \ \ \ .
	\end{equation}
	
	At iterations $r=p-2$ to $r=2$, Gr selects a vertex with maximum coverage from the largest subpath induced by removing the covered vertices. The largest subpath has at least $2r+1$ vertices, i.e.,
	
	\begin{equation}
		\label{eq:2}
		\max \{|P_{a}|,|P_{b}|\} \ge \Bigl \lceil \frac{(r+2)^2 - (2(r+1)+1)}{2} \Bigr \rceil \ge 2r+1 \ \ \ \ ,
	\end{equation}
	
	where $(r+2)^2$ is the number of uncovered vertices two iterations before, and $2(r+1)+1$ is the number of newly covered vertices at the previous iteration. Thus, the numerator is the number of uncovered vertices at the beginning of iteration $r$. This way, at iteration $r$, Gr selects a vertex that covers $2r+1$ vertices, none of which were covered in the previous iterations. Notice that ties can be broken at random. By mathematical induction, Eq. (\ref{eq:2}) holds for any integer $r\ge 2$.
	
	At iteration $r=1$, the number of previously covered vertices equals $\sum_{r=2}^{p-1}(2r+1)=p^2-4$. Namely, at this point, only four vertices are left uncovered. By direct inspection, there will be cases where only two or three new vertices can be covered at iterations $r=1$ and $r=0$, leaving the rest uncovered (See Figure \ref{fig:2}). Therefore, the constructed solution might not be a burning sequence.
	
	\begin{figure}
		\centering
		\includegraphics[scale=0.55]{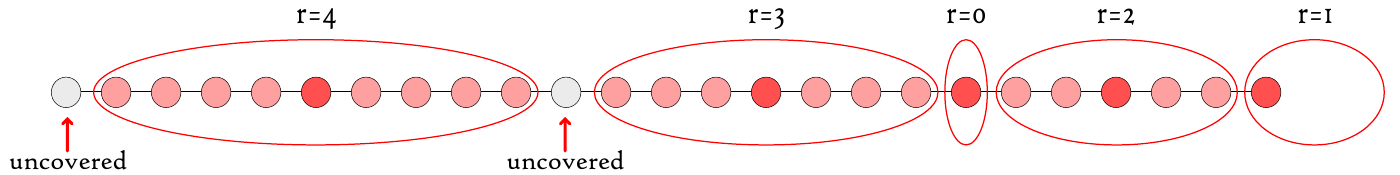}
		\caption{A non-burning sequence returned by Gr over $P_{25}$ with $p = b(P_{25})=5$. At the end, two vertices are uncovered.}
		\label{fig:2}
	\end{figure}
	
\end{proof}

\begin{corollary}
	\label{cor:1}
	Gr might not return a burning sequence for a cycle $C$.
\end{corollary}

\begin{proof}
	Since a cycle $C_n$ is a path $P_n$ with its endpoints joined, it follows that $b(C_n)=\lceil n^{1/2} \rceil$ \cite{bonato2014burning}. So, we can run Gr with $p=b(C_n)$, which is equivalent to execute Algorithm \ref{alg:bs} with $l=h=b(C_n)$. The first vertex selected by Gr maximizes the coverage. Then, removing the covered vertices induces a path $P$. By Proposition \ref{prop:3}, Gr might return a non-burning sequence of length $b(C_n)-1$ over $P$, which is concatenated to the first selected vertex, becoming a non-burning sequence of length $b(C_n)$ for $C_n$.
\end{proof}

By Proposition \ref{prop:3} and Corollary \ref{cor:1}, Gr might not seem very appealing. However, as reported in Section \ref{sec:exps}, it performed unexpectedly well over benchmark and synthetic graphs. Anyhow, we can do better by repeating Gr up to $n$ times and selecting a different initial vertex every time; we refer to this heuristic as GrP (Algorithm \ref{alg:grp}). Once the first burning sequence of length $p$ is found (line 3), exploring the remaining vertices is unnecessary. Nevertheless, the complexity of GrP embedded into a binary search (Algorithm \ref{alg:bs}) is $\mathcal{O}(n^4 \log n)$, which is impractical for sufficiently large graphs.

\begin{algorithm}
	\caption{Greedy burning plus (GrP)}\label{alg:grp}
	\begin{algorithmic}[1]
		\Require A graph $G=(V,E)$, its distance matrix $D$, and a guess $p$ on $b(G)$
		\Ensure A sequence
		\For{$v \in V$}
		\State $S \leftarrow Gr(G,p,d)$ with $v$ as the first selected vertex
		\If{$S$ burns all vertices}
		\State \Return $S$
		\EndIf
		\EndFor \\
		\Return the best found sequence
	\end{algorithmic}
\end{algorithm}
	
\section{Experimental results}
	\label{sec:exps}
This section reports the execution time and the size of the burning sequences returned by Algorithm \ref{alg:bs} using Gr (Algorithm \ref{alg:gr}) and GrP (Algorithm \ref{alg:grp}) over benchmark and synthetic graphs. Off-the-shelf optimization software implementing ILP-COV (Expressions (\ref{IP2:1}) to (\ref{IP2:7})) was used to compute optimal solutions; the upper bound $U$ being equal to the length of the best feasible solution returned by Gr and GrP. All the real-world networks were obtained from the network repository \cite{nr} and the Stanford large network dataset collection \cite{snapnets}. Gr and GrP were implemented in C++ and compiled with GNU GCC. ILP-COV was solved to optimality by Gurobi 11.0.0 \cite{gurobi} using default parameters. The execution times of Gurobi using ILP-PROP and ILP-CMCP (Algorithm \ref{alg:trans}) were larger; thus, we did not report them. Table \ref{tab:hardware} shows all experiments' hardware and software specifications. The implemented codes can be consulted at \url{https://github.com/jesgadiaz/GreedyBurning}. 

\begin{table}
	\centering
	\caption{Hardware and software setup}\label{tab:hardware}
	\begin{tabular}{@{}ll@{}}
		\toprule
		\multicolumn{2}{l}{For all experiments} \\
		\midrule
		\hspace{0.3cm} CPU &  Intel Core i5-10300H\\
		\hspace{0.3cm} CPU frequency & 2.5GHz\\
		\hspace{0.3cm} CPU cores & 4\\
		\hspace{0.3cm} RAM & 32GB\\
		\hspace{0.3cm} OS & Windows 11\\
		\midrule
		\multicolumn{2}{l}{For Gr and GrP} \\
		\midrule
		\hspace{0.3cm} Programming language & C++ \\
		\hspace{0.3cm} Compiler & GNU GCC \\
		\midrule
		\multicolumn{2}{l}{For ILP} \\
		\midrule
		\hspace{0.3cm} Solver &Gurobi 11.0.0 \\
		\hspace{0.3cm} Presolve & -1 \\
		\hspace{0.3cm} MIPGap & 0\% \\
		\hspace{0.3cm} Heuristics & 5\% \\
		\hline
	\end{tabular}
\end{table}

Table \ref{tab:smallbench} (real-world graphs with less than 6000 vertices), Table \ref{tab:largebench} (real-world graphs with order between 7057 and 88784), and Table \ref{tab:grids} (square grids) show the execution time and the length of the solutions found by Gr and GrP within a binary search (Algorithm \ref{alg:bs}). Column BFF reports the bounds used in the binary search; we used BFF algorithm to set $l=\lceil (s(S_0)+2)/3 \rceil$ and $h=s(S_0)-1$, where $s(S_0)$ is the size of the burning sequence returned by this algorithm; for reproducibility purposes, the first vertex selected by BFF is the one with the smallest label. Column BFS reports the time required to compute the distance matrix using breadth-first search (all reported times are in seconds). Column Gurobi reports the burning number computed by this solver using ILP-COV, where the upper bound $U$ is the size of the best feasible solution found by Gr and GrP.

Using ILP-COV, Gurobi corroborated all the previously reported optimal solutions and found six new ones (tvshow, government, crocodile, DD6, DD349, and DD687). According to these results, ILP-COV seems advantageous compared to the state-of-the-art ILP and CSPs \cite{garcia2022graph}. For instance, Gurobi found the optimal burning sequence of tvshow in 62 seconds using ILP-COV, while the same software could not find it in more than 48 hours using the ILP and CSPs reported by \cite{garcia2022graph}. Besides, while the largest instance solved by the aforementioned models has 5908 vertices, Gurobi solved an instance of order 11631 in 841 seconds ($\sim$14 minutes) using ILP-COV; we get an out-of-memory error for larger graphs. For the graphs with a dash symbol at column $t$ of the Gurobi column, we report the burning number found by a recent state-of-the-art exact algorithm \cite{pereira2024solving}; although this algorithm is faster than Gurobi using ILP-COV, we are reporting the running time of the latter for comparison and corroboration purposes. Regarding Gr and GrP, the former was the fastest among all the tested algorithms, requiring 3282 seconds ($\sim$55 minutes) over a graph of order 102400. A dash at Tables \ref{tab:smallbench} to \ref{tab:grids} indicates that the corresponding procedure was not executed because the memory requirements exceeded the hardware's capacity or a time limit of 10 hours was estimated to be surpassed. In all tables, optimal solutions are emphasized, and all times are in seconds; a time of 0.000 represents a time below milliseconds.


\begin{table}[h]
	\addtolength{\tabcolsep}{-4pt}
	\caption{Experimental results over real-world networks with less than 6000 vertices.}\label{tab:smallbench}
	\begin{tabular*}{\textwidth}{@{\extracolsep\fill}lllccccccccccc}
		\toprule%
		\multicolumn{3}{@{}c@{}}{Graph} & \multicolumn{2}{@{}c@{}}{Gurobi} & \multicolumn{3}{@{}c@{}}{BFF} & BFS & \multicolumn{2}{@{}c@{}}{Gr} & \multicolumn{2}{@{}c@{}}{GrP}\\
		\cmidrule{1-3}\cmidrule{4-5}\cmidrule{6-8}\cmidrule{9-9}\cmidrule{10-11}\cmidrule{12-13}
		name & $n$ & $m$ & $b(G)$ & $t$ & $l$ & $s(S_0)$ & $t$ & $t$ & size & $t$ & size & $t$ \\
		\midrule
		karate & 34 & 78 & 3 & 0.015 & 2 & 4 & 0.000 & 0.000 & \textbf{3} & 0.000 & \textbf{3}& 0.000\\ 
		chesapeake & 39 & 170 & 3 & 0.000 & 1 & 3 & 0.000 & 0.000 & \textbf{3} & 0.000 & \textbf{3}& 0.001\\ 
		dolphins & 62 & 159 & 4 & 0.000 & 2 & 6 & 0.000 & 0.000 & \textbf{4} & 0.001 & \textbf{4}& 0.004\\ 
		rt-retweet & 96 & 117 & 5 & 0.016 & 2 & 6 & 0.000 & 0.000 & \textbf{5} & 0.001 & \textbf{5}& 0.024\\ 
		polbooks & 105 & 441 & 4 & 0.016 & 2 & 5 & 0.000 & 0.000 & \textbf{4} & 0.000 & \textbf{4}& 0.018\\ 
		adjnoun & 112 & 425 & 4 & 0.016 & 2 & 5 & 0.000 & 0.000 & \textbf{4} & 0.000 & \textbf{4}& 0.032\\ 
		ia-infect-hyper & 113 & 2196 & 3 & 0.032 & 1 & 3 & 0.000 & 0.015 & \textbf{3} & 0.000 &\textbf{3} &0.018 \\ 
		C125-9 & 125 & 6963 & 3 & 0.125 & 1 & 3 & 0.000 & 0.031 & \textbf{3} & 0.000 & \textbf{3}& 0.055\\ 
		ia-enron-only & 143 & 623 & 4 & 0.031 & 2 & 5 & 0.000 & 0.000 & \textbf{4} & 0.002 & \textbf{4}& 0.046\\ 
		c-fat200-1 & 200 & 1534 & 7 & 0.044 & 3 & 7 & 0.000 & 0.015 & \textbf{7} & 0.002 & \textbf{7}& 0.307\\ 
		c-fat200-2 & 200 & 3235 & 5 & 0.029 & 2 & 5 & 0.000 & 0.031 & \textbf{5} & 0.003 & \textbf{5}&0.240 \\ 
		c-fat200-5 & 200 & 8473 & 3 & 0.016 & 1 & 3 & 0.000 & 0.062 & \textbf{3} & 0.000 & \textbf{3}& 0.105\\ 
		sphere & 258 & 1026 & 7 & 1.6 & 3 & 9 & 0.000 & 0.015 & \textbf{7} & 0.006 & \textbf{7}& 0.488\\ 
		DD244 & 291 & 822 & 7 & 0.244 & 4 & 11 & 0.000 & 0.015 & \textbf{7} & 0.007 & \textbf{7}& 0.756\\ 
		ca-netscience & 379 & 914 & 6 & 0.077 & 3 & 8 & 0.000 & 0.031 & 7 & 0.016 & \textbf{6}&0.643 \\ 
		infect-dublin & 410 & 2765 & 5 & 0.204 & 2 & 6 & 0.000 & 0.062 & \textbf{5} & 0.018 & \textbf{5}& 1.9\\ 
		c-fat500-1 & 500 & 4459 & 9 & 0.174 & 4 & 11 & 0.000 & 0.106 & \textbf{9} & 0.015 & \textbf{9}&2.7 \\ 
		c-fat500-2 & 500 & 9139 & 7 & 0.142 & 3 & 8 & 0.000 & 0.203 & \textbf{7} & 0.018 & \textbf{7}& 3.2\\ 
		c-fat500-5 & 500 & 23191 & 5 & 0.255 & 2 & 5 & 0.000 & 0.456 & \textbf{5} & 0.014 &\textbf{5} &3.9 \\ 
		bio-diseasome & 516 & 1188 & 7 & 0.174 & 5 & 13 & 0.000 & 0.046 & \textbf{7} & 0.028 & \textbf{7}&2.2 \\ 
		web-polblogs & 643 & 2280 & 5 & 0.440 & 3 & 8 & 0.000 & 0.109 & 6 & 0.049 & \textbf{5}&5.0 \\ 
		DD687 & 725 & 2600 & 7 & 4.1 & 4 & 10 & 0.000 & 0.125 & 8 & 0.053 & 8&14 \\ 
		rt-twitter-copen & 761 & 1029 & 7 & 0.852 & 3 & 9 & 0.000 & 0.093 & \textbf{7} & 0.061 &\textbf{7} &14 \\ 
		DD68 & 775 & 2093 & 9 & 2.1 & 5 & 14 & 0.000 & 0.124 & 10 & 0.059 &\textbf{9} &16 \\ 
		ia-crime-moreno & 829 & 1475 & 7 & 4.2 & 3 & 8 & 0.000 & 0.112 & \textbf{7} & 0.075 & \textbf{7}&20 \\ 
		DD199 & 841 & 1902 & 12 & 1.7 & 6 & 16 & 0.000 & 0.125 & 13 & 0.066 & \textbf{12}&13 \\ 
		soc-wiki-Vote & 889 & 2914 & 6 & 1.4 & 3 & 8 & 0.000 & 0.171 & \textbf{6} & 0.065 & \textbf{6}&16 \\ 
		DD349 & 897 & 2087 & 12 & 1.9 & 6 & 18 & 0.000 & 0.157 & 13 & 0.076 & \textbf{12}&8.6 \\ 
		DD497 & 903 & 2453 & 10 & 2.1 & 6 & 16 & 0.000 & 0.160 & 12 & 0.086 & 11&9.4 \\ 
		socfb-Reed98 & 962 & 18812 & 4 & 2.4 & 2 & 5 & 0.000 & 0.735 & \textbf{4} & 0.077 & \textbf{4}&16 \\ 
		lattice3D & 1000 & 2700 & 10 & 116 & 4 & 12 & 0.000 & 0.203 & \textbf{10} & 0.089 & \textbf{10}&34 \\ 
		bal-bin-tree-9 & 1023 & 1022 & 10 & 0.883 & 4 & 10 & 0.000 & 0.125 & \textbf{10} & 0.024 & \textbf{10}&8.2 \\ 
		delaunay-n10 & 1024 & 3056 & 9 & 5.6 & 4 & 11 & 0.000 & 0.236 & 10 & 0.072 & \textbf{9}&27 \\ 
		stufe & 1036 & 1868 & 12 & 13 & 5 & 15 & 0.000 & 0.176 & \textbf{12} & 0.076 & \textbf{12}&30 \\ 
		lattice2D & 1089 & 2112 & 13 & 22 & 7 & 19 & 0.000 & 0.187 & 14 & 0.099 &\textbf{13} &20 \\ 
		bal-ter-tree-6 & 1093 & 1092 & 7 & 0.343 & 3 & 7 & 0.000 & 0.157 & \textbf{7} & 0.022 & \textbf{7}&4.7 \\ 
		email-univ & 1133 & 5451 & 5 & 3.1 & 2 & 6 & 0.000 & 0.359 & \textbf{5} & 0.105 & \textbf{5}&29 \\ 
		econ-mahindas & 1258 & 7513 & 5 & 3.5 & 2 & 6 & 0.000 & 0.488 & \textbf{5} & 0.126 & \textbf{5}&40 \\ 
		ia-fb-messages & 1266 & 6451 & 5 & 3.7 & 2 & 6 & 0.000 & 0.461 & \textbf{5} & 0.152 & \textbf{5}&54 \\ 
		bio-yeast & 1458 & 1948 & 9 & 7.4 & 4 & 11 & 0.000 & 0.343 & \textbf{9} & 0.237 & \textbf{9}&141 \\ 
		tech-routers-rf & 2113 & 6632 & 6 & 9.4 & 3 & 8 & 0.000 & 1.0 & \textbf{6} & 0.314 &\textbf{6} &194 \\ 
		chameleon & 2277 & 31421 & 6 & 14.1& 3 & 8 & 0.000 & 3.4 & \textbf{6} & 0.444 &\textbf{6} &334 \\ 
		tvshow & 3892 & 17262 & 9 & 62 & 5 & 13 & 0.000 & 4.0 & 10 & 1.9 & \textbf{9}&3057 \\ 
		facebook & 4039 & 88234 & 4 & 9.2 & 2 & 5 & 0.000 & 14 & \textbf{4} & 0.547 & \textbf{4}& 377\\ 
		DD6 & 4152 & 10320 & 16 & 72 & 9 & 25 & 0.000 & 3.4 & 17 & 1.3 & 17& 821\\
		squirrel & 5201 & 198493 & 6 & 124 & 2 & 6 & 0.000 & 38 & \textbf{6} & 2.6 & \textbf{6}& 8853\\ 
		politician & 5908 & 41729 & 7 & 113 & 3 & 9 & 0.000 & 13 & \textbf{7} & 3.8 & \textbf{7}& 9068\\ 
		\hline
	\end{tabular*}
\end{table}



\begin{table}[h]
	\addtolength{\tabcolsep}{-4pt}
	\caption{Experimental results over real-world networks with order between 7057 and 88784.}\label{tab:largebench}
	\begin{tabular*}{\textwidth}{@{\extracolsep\fill}lllccccccccccc}
		\toprule%
		\multicolumn{3}{@{}c@{}}{Graph} & \multicolumn{2}{@{}c@{}}{Gurobi} & \multicolumn{3}{@{}c@{}}{BFF} & BFS & \multicolumn{2}{@{}c@{}}{Gr} & \multicolumn{2}{@{}c@{}}{GrP}\\
		\cmidrule{1-3}\cmidrule{4-5}\cmidrule{6-8}\cmidrule{9-9}\cmidrule{10-11}\cmidrule{12-13}
		name & $n$ & $m$ & $b(G)$ & $t$ & $l$ & $s(S_0)$ & $t$ & $t$ & size & $t$ & size & $t$ \\
		\midrule
		government & 7057 & 89455 & 6 & 178 & 3 & 7 & 0.000 & 28 & \textbf{6} & 5.3 & \textbf{6}& 14973\\ 
		crocodile & 11631 & 170918 & 6 & 841 & 3 & 7 & 0.000 & 84 & \textbf{6} & 16 &\textbf{6} &81669 \\ 
		athletes & 13866 & 86858 & 7 & - & 3 & 8 & 0.000 & 65 & \textbf{7} & 23 & -&- \\ 
		company & 14113 & 52310 & 9 & -  & 4 & 10 & 0.015 & 50 & \textbf{9} & 23 & -&- \\ 
		musae-facebook & 22470 & 171002 & 8 & -  & 4 & 11 & 0.015 & 208 & \textbf{8} & 68 &- &- \\ 
		new-sites&27917&206259&8 & - &3&9&0.031&322&\textbf{8}&95&-&-\\
		deezer-europe & 28281 & 92752 & 10 & -  & 4 & 12 & 0.031 & 208 & \textbf{10} & 93 & -& -\\ 
		gemsec-deezer(RO)&41773&125826&10 & - &4&12&0.016&463&\textbf{10}&241&-&-\\
		gemsec-deezer(HU)&47538&222887&8 & - &4&10&0.046&780&\textbf{8}&325&-&-\\
		artist&50515&819306&6 & - &3&8&0.031&2035&\textbf{6}&274&-&-\\
		gemsec-deezer(HR)&54573&498202&7 & - &3&8&0.047&1631&\textbf{7}&433&-&-\\
		soc-brightkite & 56739 & 212945 & 9 & - & 3 & 9 & 0.062 & 963 &\textbf{9}&437&-&- \\
		socfb-OR & 63392 & 816886 & 8 & - &3&9&0.051&2616&\textbf{8}&795&-&- \\
		soc-slashdot & 70068 & 358647 & 7 & - &3&8&0.031&1718&\textbf{7}&1401& -&-\\
		soc-BlogCatalog & 88784 & 2093195 & 5 & - &2&6&3.9&8600&\textbf{5}&3315&-&- \\
		\hline
	\end{tabular*}
\end{table}



\begin{table}[h]
	\addtolength{\tabcolsep}{-4pt}
	\caption{Experimental results over square grids.}\label{tab:grids}
	\begin{tabular*}{\textwidth}{@{\extracolsep\fill}lllccccccccccc}
		\toprule%
		\multicolumn{3}{@{}c@{}}{Graph} & \multicolumn{2}{@{}c@{}}{Gurobi} & \multicolumn{3}{@{}c@{}}{BFF} & BFS & \multicolumn{2}{@{}c@{}}{Gr} & \multicolumn{2}{@{}c@{}}{GrP}\\
		\cmidrule{1-3}\cmidrule{4-5}\cmidrule{6-8}\cmidrule{9-9}\cmidrule{10-11}\cmidrule{12-13}
		name & $n$ & $m$ & $b(G)$ & $t$ & $l$ & $s(S_0)$ & $t$ & $t$ & size & $t$ & size & $t$ \\
		\midrule
		10x10 & 100 & 180 & 6 & 0.099 & 3 & 8 & 0.000 & 0.001 & 7 & 0.001 & \textbf{6}&0.023 \\ 
		20x20 & 400 & 760 & 10 & 1.6 & 5 & 13 & 0.000 & 0.027 & 11 & 0.018 & \textbf{10}&1.7 \\ 
		30x30 & 900 & 1740 & 12 & 46 & 6 & 17 & 0.000 & 0.14 & 13 & 0.079 & \textbf{12}&13 \\ 
		40x40 & 1600 & 3120 & 15 & 58 & 7 & 20 & 0.000 & 0.45 & 16 & 0.202 & \textbf{15}& 93\\ 
		50x50 & 2500 & 4900 & 17 & 302 & 8 & 23 & 0.000 & 1.1 & 18 & 0.469 & \textbf{17}& 415\\ 
		60x60 & 3600 & 7080 & 19 & 1107 & 9 & 26 & 0.011 & 2.2 & 21 & 0.873 & \textbf{19}&1219 \\ 
		70x70 & 4900 & 9660 & 21 & 20073 & 10 & 29 & 0.000 & 4 & 22 & 1.5 & \textbf{21}& 3068\\ 
		80x80 & 6400 & 12640 & ? & - & 11 & 31 & 0.000 & 7 & 25 & 3.2 & 24&8241 \\ 
		90x90 & 8100 & 16020 & ? & -  & 12 & 34 & 0.000 & 11 & 28 & 5.0 & 25&1756 \\ 
		100x100 & 10000 & 19800 & ? & -  & 12 & 36 & 0.013 & 17 & 28 & 7.4 & 27&22301\\
		110x110 & 12100 & 23980 & ? & -  & 13 & 38 & 0.013 & 25 & 30 & 10 & -&- \\ 
		120x120 & 14400 & 28560 & ? & -  & 14 & 41 & 0.015 & 36 & 32 & 15 & -&- \\ 
		130x130 & 16900 & 33540 & ? & -  & 15 & 43 & 0.015 & 50 & 34 & 20 & -&- \\ 
		140x140 & 19600 & 38920 & ? & -  & 16 & 46 & 0.046 & 69 & 35 & 27 & -&- \\ 
		150x150 & 22500 & 44700 & ? & -  & 16 & 48 & 0.046 & 93 & 37 & 35 & -&- \\ 
		160x160 & 25600 & 50880 & ? & -  & 17 & 51 & 0.062 & 123 & 39 & 44 & -&- \\ 
		170x170 & 28900 & 57460 & ? & -  & 18 & 53 & 0.063 & 161 & 41 & 56 & -&- \\ 
		180x180 & 32400 & 64440 & ? & -  & 19 & 55 & 0.063 & 208 & 42 & 70 & -&- \\ 
		190x190 & 36100 & 71820 & ? & -  & 19 & 57 & 0.094 & 259 & 44 & 86 & -&- \\ 
		200x200 & 40000 & 79600 & ? & -  & 20 & 60 & 0.141 & 322 & 45 & 105 & -&- \\ 
		210x210 & 44100 & 87780 & ? & -  & 20 & 60 & 0.126 & 397 & 47 & 126 & -&- \\ 
		220x220 & 48400 & 96360 & ? & -  & 21 & 63 & 0.157 & 487 & 48 & 182 & -&- \\ 
		230x230 & 52900 & 105340 & ? & -  & 22 & 64 & 0.181 & 591 & 50 & 218 & -&- \\ 
		240x240 & 57600 & 114720 & ? & -  & 23 & 67 & 0.188 & 713 & 51 & 257 & -&- \\ 
		250x250 & 62500 & 124500 & ? & -  & 23 & 67 & 0.345 & 838 & 52 & 301 & -& -\\ 
		260x260 & 67600 & 134680 & ? & -  & 23 & 69 & 0.219 & 1019 & 54 & 353 & -&- \\ 
		270x270 & 72900 & 145260 & ? & -  & 24 & 70 & 0.464 & 1174 & 55 & 404 & -& -\\ 
		280x280 & 78400 & 156240 & ? & -  & 25 & 73 & 0.456 & 1388 & 57 & 478 &- &- \\ 
		290x290 & 84100 & 167620 & ? & -  & 25 & 73 & 5.2 & 1627 & 58 & 575 & -&- \\ 
		300x300 & 90000 & 179400 & ? & -  & 25 & 74 & 18.6 & 1954 & 59 & 1010 & -&- \\ 
		310x310 & 96100 & 191580 & ? & -  & 26 & 76 & 53.4 & 2306 & 60 & 2101 & -&- \\ 
		320x320 & 102400 & 204160 & ? & -  & 26 & 78 & 35 & 2687 & 62 & 3282 & -&- \\ 
		\hline
	\end{tabular*}
\end{table}
	
\section{Discussion}
	
GBP is an NP-hard minimization problem with application to information spreading problems. Some heuristics and metaheuristics have been proposed to find feasible solutions. However, research on more straightforward greedy approaches wasn't fully addressed. This paper introduces a simple deterministic greedy heuristic, Gr (Algorithm \ref{alg:gr}), tailored for the CMCP, a problem to which GBP is reduced if $b(G)$ is known in advance; the issue of not knowing this value is tackled by using binary search (Algorithm \ref{alg:bs}). Although Gr cannot guarantee to find optimal solutions for paths and cycles for which exact $\mathcal{O}(n)$ algorithms exist, it performed unexpectedly well over real-world and synthetic graphs. Namely, for a benchmark dataset of real-world networks with order between 34 and 5908, Gr found 37 optimal solutions out of 47. Its \textit{plus} version, GrP, found 44 out of 47 optimal solutions (See Table \ref{tab:smallbench}). Regarding larger real-world networks, Gr found 15 out of 15 optimal solutions; for most of these graph instances, GrP was very inefficient (See Table \ref{tab:largebench}). Regarding synthetic graphs, square grids seem challenging to solve because they have a relatively large burning number. For these instances, although Gr could not find any optimal solution, GrP could. Nevertheless, the maximum execution time for Gr was 3282 seconds ($\sim$55 minutes) over a grid of order 102400, while GrP required 22301 seconds ($\sim$6.2 hours) over a grid of order 10000. Gurobi became unpractical on grids of order at least 4900 (See Table \ref{tab:grids}).

Using Gurobi for solving ILP-COV, all the previously known optimal solutions were corroborated, and six new ones were found (tvshow, government, crocodile, DD6, DD349, and DD687); ILP-COV showed a better performance than the mathematical formulations for GBP reported by \cite{garcia2022graph} (See Tables \ref{tab:smallbench} to \ref{tab:grids}).


\section{Conclusions}

This paper shows how GBP can be reduced to a series of CMCPs (See Proposition \ref{prop:red}, Algorithm \ref{alg:trans}, and Expressions (\ref{IP:1}) to (\ref{IP:4})). Through experimental results, we show that a deterministic greedy $(1/2)$-approximation algorithm for CMCP (Algorithm \ref{alg:grc}) can be used to find feasible solutions for GBP. Although the deterministic greedy heuristic for GBP, Gr (Algorithm \ref{alg:gr}), might not find optimal solutions of paths and cycles, it might have interesting theoretical properties over other graph families; we leave that research for future work. Nevertheless, the proposed heuristic performed unexpectedly well over real-world and synthetic graphs, finding most of the optimal and best-known solutions in reasonable amounts of time. Finally, the relationship between the GBP and the CMCP might be further exploited in the future, leading to more efficient algorithms.

	\bibliographystyle{unsrt}
	\bibliography{references}  

		
		
		
		

\end{document}